\documentstyle[epsf,aps]{revtex}

\begin{document}
\draft
\title {Vacuum Induced Coherences in Radiatively Coupled Multilevel Systems}

\author {G. S. Agarwal\footnote{also at Jawaharlal Nehru Center for Advanced Scientific Research, Bangalore, India} and Anil K. Patnaik}
\address {Physical Research Laboratory, Navrangpura, Ahmedabad-380 009, India}
\date{\today}

\maketitle 
\begin{abstract} 
We show that radiative coupling between two multilevel atoms
having near-degenerate states can produce new interference effects
in spontaneous emission.  We explicitly demonstrate this possibility by
considering two identical V systems each having a pair of transition 
dipole matrix elements which are orthogonal to each other. We  discuss 
in detail the origin of the new interference terms and their consequences. 
Such terms lead to the evolution of certain coherences and excitations 
which would not occur otherwise. The special
choice of the orientation of the transition dipole matrix elements enables 
us to illustrate the significance of
vacuum induced coherence in multi-atom multilevel systems.
These coherences can be significant in energy transfer studies.
\end{abstract}


\section{Introduction}
A multilevel atom having closely lying energy states (energy 
separation of the order of the natural linewidth) can show interferences
in the decays from those closely lying levels to a common 
ground state. This is due to the fact that,
both the decay channels are coupled via the same continuum of the 
vacuum, creating the interfering path ways. The 
resulting coherence in the system is known as {\em vacuum induced 
coherence} (VIC). Occurrence of this coherence requires a stringent but
achievable condition - i.e. the transition dipole matrix elements involving 
the decay processes should be {\em non-orthogonal} \cite{gsabook}. 
The manifestation of VIC in atomic systems has given rise to a myriad 
of fascinating phenomena [1-17]. All these studies deal with a 
{\em single} multilevel atom or equivalently with an ensemble of 
non-interacting 
multilevel atoms (e.g. very low density atomic gas systems). However, 
VIC in {\em coupled} atomic systems has remained unexplored. 
In this paper, we consider the role of VIC in {\em two radiatively
coupled multilevel atoms}. 

We start by recalling some of the consequences of VIC in a single atom. 
It was first shown by Agarwal \cite{gsabook} that population gets
trapped in degenerate excited states of an atom due to interference in decays 
channels. Recently, there has been renewed  interest in this subject 
particularly in the context of coherently driven systems 
\cite{cpt-card,cpt-swain,cpt-menon,cpt-lineshape-menon}. Harris and Imamo\u{g}lu
were
the first to discover the possibility of achieving lasing without population
inversion in systems where two excited states were coupled to a common
continuum \cite{lwi-harris} (See also \cite{lwi-swain,lwi-knight}). 
It has also been been observed that narrowing of spontaneous emission
can be obtained by making use of the VIC \cite{cpt-swain,sub-nat-keittel}. 
Quantum beat has been observed
in spontaneous emission which showed pronounced beat structures determined
by the energy separation of the closely lying states 
\cite{qbeat-hegerfeldt,qbeat-patnaik}. The VIC also leads to cancellation
of spontaneous emission 
\cite{quench-zhu,quench-scully,quench-berman}. Zhu and coworkers
have experimentally demonstrated the quenching of spontaneous emission 
in sodium dimers \cite{quench-zhu}. Further
Scully, Zhu and coworkers proposed many schemes with different configurations
demonstrating the possibility of obtaining quenching of spontaneous emission
\cite{quench-scully}. It was also reported that in the presence of VIC, 
the resonance fluorescence \cite{phase-dep-martinez,phase-dep-knight} and
other spectral line shapes \cite{cpt-lineshape-menon} become sensitive to the 
phase of the control laser. Knight and coworkers \cite{phase-dep-knight} have
demonstrated the possibility of controlling spontaneous emission by varying
the relative phase of two control lasers in a four-level system. The 
question of requirement of non-orthogonal dipole moments has been addressed
and alternative possibilities have been suggested 
\cite{qbeat-patnaik,quench-zhu,quench-berman,aniso-gsa}. 

All the above works [2-17] correspond to a single atom system, or equivalently 
an ensemble of non-interacting atoms where the average inter-atomic
distance is much larger compared to the wavelength of the emitted radiation.  
However, when the inter-atomic distance becomes comparable to the wavelength,
the dipole-dipole (dd) coupling between the atoms via vacuum gives rise to 
collective effects. Our usage of dd interaction should be understood
in the sense of retarded (and complex) dipole-dipole interaction.
The classic example is Dicke superradiance 
\cite{super-R-haroche}, where the atoms in their excited state decay much 
faster compared to that of the single atom case. The collective effects 
in atoms have been extensively studied [18-30]. Recently
experiments have been reported to observe collective behavior
with two identical atoms \cite{super-R-brewer,super-R-martini}. 
The dd interaction has been shown to produce two-photon resonance 
\cite{dd-2photon} and frequency shifts in emission \cite{dd-freq-shift}. 
The energy exchange between two coupled systems is discussed in 
\cite{dd-energy-exch}. Many interesting features of dd interaction in the 
context of atoms interacting with a squeezed vacuum \cite{dd-squeez}, and 
inside bandgap materials \cite{dd-bandgap} have been reported. Mayer and 
Yeoman \cite{2atm-laser-mayer} have considered two-atom laser in the presence 
of the atom-atom interaction. Quantum jump from two dipole interacting 
V-systems giving rise to new fluorescence periods has been reported by 
Hegerfeldt and coworkers \cite{dd-qjump-hegerfeldt}. Meystre and
coworkers \cite{dd-dark-meystre} found that the dd interaction leads
to the occurrence of dark states in the fluorescence 
of two moving atoms \cite{dd-collision-harris}. Finally note that the 
local field effects in a dense media are also a consequence of 
dd interaction \cite{dd-bowden}.
All the dd interaction related effects can be understood as
due to the exchange of virtual photons between the atoms. 
Most of the existing literature concerns two-level atoms.

In this paper, we consider two identical V-systems having two closely lying 
excited states (as shown in Fig.\ 1). The two atoms get coupled 
by the exchange of radiation. We would specifically show how the radiative
coupling in multilevel systems can lead to a population transfer from
$|1_A\rangle$ to $|2_B\rangle$ even if the corresponding dipole matrix 
elements are orthogonal.

\begin{figure}
\epsfxsize 10cm
\centerline{
\epsfbox{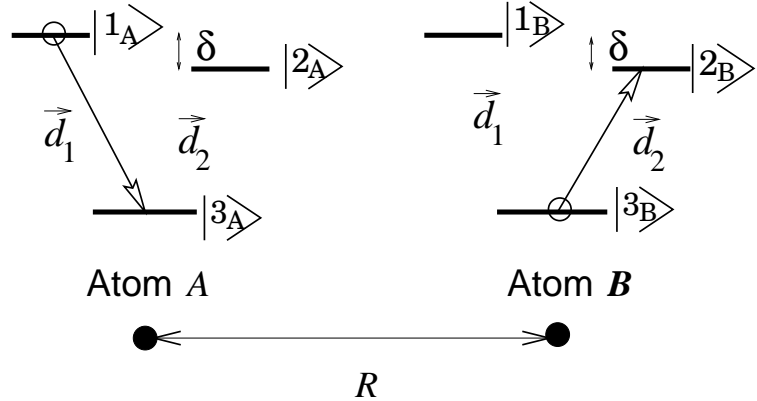}
}
\caption{
The two identical $V$ systems under consideration. 
The distance between the two atoms is $R$. The transition dipole matrix 
elements $\vec{d}_1$ and $\vec{d}_2$ are chosen to be orthogonal with each
other. The energy separation between the excited states is $\hbar\delta$.
We discuss the situation where initially atom $A$ is in
excited state $|1_A\rangle$ and atom $B$ is in ground state $|3_B\rangle$.
Possibility of dd interaction induced excitation in the 
$|3_B\rangle \rightarrow |2_B\rangle$ transition is indicated.
}
\end{figure}

	The organization of the paper is the following: In Sec. II we
derive the equations for the dynamical evolution of the two V-systems
in master equation formalism. In Sec. III we interpret the terms appearing 
in the master equation and discuss the dependence of these terms
on the geometry of the atoms in detail. In Sec.IV we present the numerical 
results on the evolution of the excited state coherence and the excitation 
probabilities. In Sec. V we present results for the case of magnetic 
degeneracies and we also discuss how the new coherence effect can be monitored.
In Sec.VI we present the concluding remarks.

\section{Dynamical Evolution of Two V-systems}

	We consider [Fig.\ 1] two identical V-systems (say A and B) in free 
space, having two near-degenerate excited states $|1_\mu\rangle$ and 
$|2_\mu\rangle$ ($\mu =A, B$) with the level separation $\hbar\delta$.
The ground states of the atoms are represented by $|3_\mu\rangle$. 
Let $\omega_1$ and $\omega_2$ be the atomic frequencies corresponding to 
$|1_\mu\rangle \leftrightarrow |3_\mu\rangle$ and
$|2_\mu\rangle \leftrightarrow |3_\mu\rangle$ transitions respectively.
Let the position vectors of the atoms be $\vec{x}_A$ and $\vec{x}_B$. 
Both the atoms couple with the vacuum field, which is given by
\begin{equation}
\vec{E}_{v} (\vec{x}) =
\vec{E}^{(+)} (\vec{x}) +\vec{E}^{(-)} (\vec{x});
\end{equation}
where $\vec{E}^{(+)}~(\vec{E}^{(-)})$ represents the positive (negative) frequency
part the vacuum field at $\vec{x}$ and is defined as
\begin{equation}
\vec{E}^{(+)}(\vec{x}) = \sum_{k,s} 
i \left(\frac{2\pi\hbar \omega_k}{V}\right)^{1/2} 
\hat{a}_{ks} \hat{\epsilon}_{ks} e^{i\vec{k}\cdot\vec{x}},
\end{equation}
and $\vec{E}^{(-)}$ is the Hermitian conjugate of $\vec{E}^{(+)}$. 
Here $\hat{a}_{ks}~(\hat{a}_{ks}^{\dag})$ represent the annihilation (creation)
operator of a field mode having propagation vector $\vec{k}$ and
polarization $s$; $\omega_k (= kc)$ represents the angular frequency 
corresponding to the $k$-mode. The corresponding unit polarization vector 
is denoted by $\hat{\epsilon}_{ks}$. 

In what follows we consider the following physical process:
Initially atom $A$ is taken to be in excited state 
$|1_A\rangle$ and the second atom in ground state $|3_B\rangle$. To highlight
the effect of new interference terms, we specifically consider the case 
when the two transition dipole matrix elements $\vec{d}_1$ 
and $\vec{d}_2$ are orthogonal to each other. In the absence of atom $B$, 
the VIC cannot be created in the excited 
states of atom $A$ because the transition dipole matrix elements $\vec{d}_1$
and $\vec{d}_2$ are orthogonal. However, the radiative coupling can lead to 
evolution of the excited state coherences. We will also show a manifestation 
of this coherence in the dynamical evolution of the atomic population.

The total Hamiltonian for the atoms and the field system is given by
\begin{equation}
H = H_{a} + H_{f} + H_{I},
\label{hamiltonian}
\end{equation}
where the unperturbed Hamiltonian for the atoms and field are
\begin{equation}
H_{a} = \hbar \sum_{\mu=A,B} \left(\omega_1 \hat{\alpha}_\mu^\dag
\hat{\alpha}_\mu + \omega_2 \hat{\beta}_\mu^\dag \hat{\beta}_\mu \right),
\label{atom-H}
\end{equation}
\begin{equation}
H_{f} = \sum_{ks} \hbar \omega_{ks} 
\hat{a}_{ks}^\dag \hat{a}_{ks},
\label{field-H}
\end{equation}
and the interaction Hamiltonian is
\begin{eqnarray}
\label{inter-H}
H_{I} &=& - \vec{d}.\vec{E}_{v} \\ \nonumber
&=& -\sum_{j=1}^2 \left[ \vec{d}_{j3}^{~(\mu)}
\cdot\vec{E}_v (\vec{x}_\mu)+ h.c.\right] ~ ~(\mu=A,~B).
\end{eqnarray}
The atomic transition operators introduced in (\ref{atom-H}), are given by
\begin{eqnarray}
\hat{\alpha}_\mu \equiv |3_\mu\rangle\langle 1_\mu|,~~
\hat{\beta}_\mu \equiv |3_\mu\rangle\langle 2_\mu|,
\nonumber \\
\hat{\alpha}_\mu^{\dag} \equiv |1_\mu\rangle\langle 3_\mu|,~~
\hat{\beta}_\mu^{\dag} \equiv |2_\mu\rangle\langle 3_\mu|.
\end{eqnarray}
Thus $\hat{\alpha}_\mu, \hat{\alpha}_\mu^{\dag}$ ($\hat{\beta}_\mu,
\hat{\beta}_\mu^{\dag}$) represent the atomic lowering and raising operators 
respectively corresponding to the $|1_\mu\rangle \leftrightarrow 
|3_\mu\rangle$ ($|2_\mu\rangle \leftrightarrow |3_\mu\rangle$) transition. 
The dipole matrix elements are represented by 
\begin{equation}
\vec{d}_{j3}^{~(\mu)} = \vec{d}_j^{~(\mu)} |j_\mu\rangle\langle3_\mu |.
\label{d-matrix}
\end{equation}
For simplicity we consider a situation where the transition dipole matrix
elements of atom $A$ are parallel to the transition dipole matrix
elements of atom $B$
\begin{equation}
\vec{d}_{13}^{~(A)}~ ||~ \vec{d}_{13}^{~(B)}~~ {\rm and}~~ 
\vec{d}_{23}^{~(A)}~ ||~ \vec{d}_{23}^{~(B)}.
\label{parallel}
\end{equation}
We also assume that 
\begin{equation}
\vec{d}_{13}^{~(A)} \cdot \vec{d}_{23}^{~(A)^*} = 0.
\end{equation}
Thus the index $\mu$ in the right
hand side of Eq.\ (\ref{d-matrix}) can be dropped and the dipole matrix elements
can be rewritten as
\begin{equation}
\vec{d}_{j3}^{~(\mu)} = \vec{d}_j |j_\mu\rangle\langle3_\mu |.
\label{final-d}
\end{equation}
Here we note that $\vec{d}_j$, in general, can be complex (see for example 
Sec.V). Using Eq.\ (\ref{final-d}), the interaction Hamiltonian in (\ref{inter-H})
reduces to
\begin{equation}
H_I = -\sum_{\mu=A,B} \left[ \left( 
\vec{d}_1 \hat{\alpha}_\mu^\dag 
+ \vec{d}_2 \hat{\beta}_\mu^\dag 
\right) 
\cdot \vec{E}_v (\vec{x}_\mu)
+ h.c.\right]. 
\label{int-H-sch}
\end{equation}
We work in the interaction picture by transforming (\ref{int-H-sch})
\begin{eqnarray}
H_I (t) &=& e^{\frac{i}{\hbar} (H_{a} + H_{f})t} H_{I} 
e^{-\frac{i}{\hbar} (H_{a} + H_{f})t}
\nonumber \\
&=& -\sum_{\mu=A,B} \left[ 
\left( \vec{d}_1 \hat{\alpha}_\mu^\dag e^{i\omega_1 t}
+ \vec{d}_2 \hat{\beta}_\mu^\dag e^{i\omega_2 t}\right) 
\cdot \left(\vec{E}^{(+)}(\vec{x}_\mu,t) 
+ \vec{E}^{(-)} (\vec{x}_\mu,t)\right)
+ h.c.\right];
\end{eqnarray}
where, 
\begin{equation} 
\vec{E}^{(+)} (\vec{x}_\mu, t) =  
\sum_{k,s} i \left(\frac{2\pi\hbar \omega_k}{V}\right)^{1/2}
\hat{a}_{ks} \hat{\epsilon}_{ks} e^{i(\vec{k}\cdot\vec{x}_\mu -\omega_k t)}
~~{\rm and}~~ 
\vec{E}^{(-)}(\vec{x}_\mu,t) = [\vec{E}^{(+)}(\vec{x}_\mu,t)]^{\dag}.
\label{Ep-Em}
\end{equation}

Let the density operator of the combined atom-field system in the 
interaction picture be represented by $\rho(t)$ which satisfies the
Liouville equation of motion
\begin{equation}
\frac{\partial\rho}{\partial t} = -\frac{i}{\hbar} \left[ H_I(t), \rho 
\right]. 
\label{first-L}
\end{equation}
To derive useful information about the evolution of the atomic system, we
derive a master equation for the reduced atomic operator by 
using the standard projection operator techniques \cite{gsabook}. We make 
certain approximations in deriving the master equation:
(a) at $t=0$, $\rho(0)$ can be factorized
into a product of atom [$\rho_a$] and field [$\rho_f$] density operators, i.e. 
$\rho(0) \equiv \rho_a (0) \rho_f (0)$. Furthermore, we invoke (b) the Born
approximation and (c) the Markoff approximation. The Born approximation
depends on the weak coupling between the vacuum and the atoms. The Markoff
approximation holds because the vacuum has fairly flat density of states.
Using the above approximations and
tracing over the field states, the density matrix equation for the atoms
becomes
\begin{equation}
\frac{\partial{\rho_a}}{\partial t} = -\frac{1}{\hbar^2}\lim_{t\rightarrow\infty} 
\int_0^t d\tau {\rm Tr}_f \left[ H_I(t), 
[H_I(t-\tau), \rho_f(0){\rho_a}] \right].
\label{master-int}
\end{equation}
The trace over the field operators inside the integral 
in Eq.\ (\ref{master-int}) is calculated using the following relations: 
\begin{eqnarray}
{\rm Tr}_f (\rho_f a^\dag_{ks} a_{k^\prime s^\prime}) = 0,
~{\rm Tr}_f (\rho_f a_{ks} a^\dag_{k^\prime s^\prime}) = 
\delta_{kk^\prime}\delta_{ss^\prime},
\nonumber \\
{\rm Tr}_f (\rho_f a_{ks} a_{k^\prime s^\prime}) =
{\rm Tr}_f (\rho_f a^\dag_{ks} a^\dag_{k^\prime s^\prime}) = 0.
\label{fld-op}
\end{eqnarray}
One also uses the rotating wave approximation (RWA) to drop the anti-resonant
terms like $\hat{\alpha}_{\mu} \hat{\alpha}_{\mu}$, 
$\hat{\alpha}_{\mu}^\dag \hat{\alpha}_{\mu}^\dag$,
$\hat{\beta}_{\mu} \hat{\beta}_{\mu}$ and
$\hat{\beta}_{\mu}^\dag \hat{\beta}_{\mu}^\dag$ in (\ref{master-int}).
Using the above conditions and carrying out a long algebra, we obtain the 
master equation for the atomic density operator
\begin{eqnarray}
\frac{\partial\rho}{\partial t} =
&-&\left[ \left\{
\gamma_1 \sum_{\mu =A,B} \left (\hat{\alpha}_\mu^\dag \hat{\alpha}_\mu \rho
-2\hat{\alpha}_\mu \rho \hat{\alpha}_\mu^\dag 
+ \rho \hat{\alpha}_\mu^\dag \hat{\alpha}_\mu \right) 
\right\}
+ 1\rightarrow 2,~ \alpha \rightarrow \beta\right]
\nonumber \\
&-&\left[ \Gamma_1
\left\{ 
(\hat{\alpha}_A^\dag \hat{\alpha}_B \rho
- 2\hat{\alpha}_B \rho \hat{\alpha}_A^\dag
+\rho \hat{\alpha}_A^\dag \hat{\alpha}_B) +~ {\rm h.c.} 
\right\} + 1 \rightarrow 2, ~ \alpha \rightarrow \beta \right]
\nonumber \\
&+& \left[ \left\{
i\Omega_1 [\hat{\alpha}_A^\dag\hat{\alpha}_B, \rho] + h.c.
\right\} + 1 \rightarrow 2, ~ \alpha \rightarrow \beta
\right]
\nonumber \\
&-&\left[ \left\{
\Gamma_{vc} \left(\hat{\beta}_A^\dag \hat{\alpha}_B \rho
- 2 \hat{\alpha}_B \rho \hat{\beta}_A^\dag
+ \rho \hat{\beta}_A^\dag \hat{\alpha}_B 
\right)e^{-i\delta t} + h.c. 
\right\}
+ A \leftrightarrow B
\right]
\nonumber \\
&+& \left[ \left\{
i \Omega_{vc} [\hat{\beta}_A^\dag \hat{\alpha}_B, \rho] e^{-i\delta t} 
+ h.c. \right\}
+ A \leftrightarrow B \right];
\label{master}
\end{eqnarray}
where,
\begin{eqnarray}
\label{coeff-master}
\gamma_i &=& \sum_{k,s} 
\left( \frac{2\pi\omega_k}{\hbar V} \right)
\pi\delta(\omega_0 -\omega_k) |\vec{d}_i\cdot\hat{\epsilon}_{ks}|^2,
\nonumber \\
\Gamma_i &=& \sum_{k,s}
\left( \frac{2\pi\omega_k}{\hbar V} \right)
\pi\delta(\omega_0 -\omega_k) |\vec{d}_i\cdot\hat{\epsilon}_{ks}|^2 
e^{i\vec{k}\cdot\vec{R}},
\nonumber \\
\Omega_i &=& \sum_{k,s}
\left( \frac{2\pi\omega_k}{\hbar V} \right)
\left( \frac{1}{\omega_0 -\omega_k}
- \frac{1}{\omega_0 +\omega_k} \right)
|\vec{d}_i\cdot\hat{\epsilon}_{ks}|^2 
e^{i\vec{k}\cdot\vec{R}},
\\ \nonumber
\Gamma_{vc} &=& \sum_{k,s}
\left( \frac{2\pi\omega_k}{\hbar V} \right)
\pi\delta(\omega_0 -\omega_k) 
(\vec{d}_2\cdot\hat{\epsilon}_{ks})(\vec{d}_1\cdot\hat{\epsilon}_{ks})^*
e^{i\vec{k}\cdot\vec{R}},
\nonumber \\
\Omega_{vc} &=& \sum_{k,s}
\left( \frac{2\pi\omega_k}{\hbar V} \right)
\left( \frac{1}{\omega_0 -\omega_k}
- \frac{1}{\omega_0 +\omega_k} \right)
(\vec{d}_2\cdot\hat{\epsilon}_{ks})(\vec{d}_1\cdot\hat{\epsilon}_{ks})^*
e^{i\vec{k}\cdot\vec{R}}.
\nonumber
\end{eqnarray}
Here, $\vec{R} = \vec{x}_A -\vec{x}_B$. Since the states $|1_\mu\rangle$
and $|2_\mu\rangle$ are closely lying, we have set
$\omega_1 \cong \omega_2 \equiv \omega_0$. The suffix $a$ of $\rho_a$
in Eq.\ (\ref{master}) has been dropped for brevity.
We have also dropped the Lamb shift terms associated with the spontaneous 
emission of the individual atoms. 
The summation over the polarization components is evaluated using the relation
\begin{equation}
\sum_s (\hat{\epsilon}_{ks})_\mu (\hat{\epsilon}_{ks}^*)_\nu \equiv 
\delta_{\mu\nu} - \hat{k}_\mu \hat{k}_\nu,
\end{equation}
where $\hat{k}_l$ represents the direction cosine of $\vec{k}/|\vec{k}|$
along the $l^{th}$ Cartesian axis.
Taking the limit $V\rightarrow \infty$ and replacing the summation over $k$
by integration over the continuum of the field modes, the terms in 
(\ref{coeff-master}) become
\begin{eqnarray}
\label{coeff}
\gamma_i = \frac{2|\vec{d}_i|^2}{3\hbar}
\left( \frac{\omega_0}{c} \right)^3 &,&
\nonumber \\
\Gamma_i = \frac{1}{\hbar} ~(\vec{d}_i \cdot 
{\rm Im}~\stackrel\Rightarrow{\chi} \cdot \vec{d}_i^*)&,&~~
\Omega_i = \frac{1}{\hbar} ~(\vec{d}_i \cdot 
{\rm Re}~\stackrel\Rightarrow{\chi} \cdot \vec{d}_i^*),
\\ \nonumber 
\Gamma_{vc} = \frac{1}{\hbar} ~(\vec{d}_2 \cdot 
{\rm Im}~\stackrel\Rightarrow{\chi} \cdot \vec{d}_1^*)&,&~~
\Omega_{vc} = \frac{1}{\hbar} ~(\vec{d}_2 \cdot 
{\rm Re}~\stackrel\Rightarrow{\chi} \cdot \vec{d}_1^*);
\end{eqnarray}
where $\stackrel\Rightarrow{\chi}$ is a tensor whose components
are given by
\begin{eqnarray}
\label{chi}
\chi_{\mu\nu} (\vec{x}_A,\vec{x}_B,\omega_0)
&\equiv& \left( k_0^2 \delta_{\mu\nu} + 
\frac{\partial^2}{\partial {x_A}_\mu \partial {x_A}_\nu} \right)
\frac{e^{ik_0R}}{R}~~~~~(k_0 = \omega_0/c)
\nonumber \\ 
&\equiv& \left[ \delta_{\mu\nu} 
\left( \frac{k_0^2}{R} +\frac{ik_0}{R^2} - \frac{1}{R^3} \right)
- R_\mu R_\nu 
\left( \frac{k_0^2}{R^3} +\frac{3i k_0}{R^4} - \frac{3}{R^5} \right)
\right] e^{ik_0 R}; R=|\vec{x}_A -\vec{x}_B|.
\end{eqnarray}

\section{Interpretation of different terms in the master equation (18)}

The $2\gamma_i$ in Eq.\ (\ref{coeff}) represents the single atom 
spontaneous decay rate from the state $|i\rangle$ to the state $|3\rangle$.
Rest of the coefficients in (\ref{master}) are related to the coupling 
between the two $V$-systems. This coupling is produced by the 
exchange of a photon between the two systems. The dipole-dipole
interaction manifests itself through the tensor $\stackrel\Rightarrow{\chi}$
defined by Eq.\ (\ref{chi}). The tensor 
$\stackrel\Rightarrow{\chi}_{\alpha\beta} 
(\vec{x}_A, \vec{x}_B, \omega)$ has the following meaning: It 
represents the $\alpha^{th}$ component of the electric field at the
point $\vec{x}_A$, produced by an oscillating dipole of unit strength
along the direction $\beta$ and located at the point $\vec{x}_B$ 
\cite{born-wolf}. In the 
limit $c \rightarrow \infty$ ($k_0 \rightarrow 0$), it reduces to the 
static dipole-dipole interaction. $\Gamma_i$ and $\Omega_i$ 
represent the dd couplings which are related to the decay and 
level shift of the collective atomic states. These coefficients couple a pair 
of parallel dipoles 
and are well known \cite{gsabook}, particularly in the context of collective
effects in two level atoms. {\em The new coherence
terms $\Gamma_{vc}$ and $\Omega_{vc}$ are the dipole-dipole cross coupling
coefficients, which couple a pair of orthogonal dipoles}. 
The meaning of such terms can be clearly understood by calculating 
the evolution of the population in the state $|3_A,2_B\rangle$, given the 
initial condition $|1_A,3_B\rangle$. From the master equation (\ref{master}),
one can show that at $t\approx 0$,
\begin{equation}
\frac{\partial}{\partial t} 
\langle 1_A, 3_B |\rho |3_A, 2_B\rangle 
= -(\Gamma_{vc}^* + i\Omega_{vc}^*) e^{i\delta t}, 
\end{equation}
and hence
\begin{eqnarray}
\frac{\partial}{\partial t} 
\langle 3_A, 2_B |\rho |3_A, 2_B\rangle 
&=&-(\Gamma_{vc} - i\Omega_{vc}) e^{-i\delta t} 
\langle 1_A, 3_B |\rho |3_A, 2_B\rangle 
\nonumber \\
& &-(\Gamma_{vc}^* + i\Omega_{vc}^*) e^{i\delta t}
\langle 3_A, 2_B |\rho |1_A, 3_B\rangle
\nonumber \\
&=& -2 |\Gamma_{vc} - i \Omega_{vc}|^2 \left( 
\frac{\sin \delta t}{\delta} \right).
\end{eqnarray}
Thus to the lowest order in $\Gamma_{vc}$ and $\Omega_{vc}$, we obtain
\begin{equation}
\langle 3_A, 2_B |\rho |3_A, 2_B\rangle \approx 
4 |\Gamma_{vc} + i\Omega_{vc}|^2 
\left( \frac{\sin^2(\delta t/2)}{\delta^2}\right).
\label{den-elmnt}
\end{equation}
Therefore the radiative process, in which atom $A$ in the excited state 
$|1_A\rangle$ loses its excitation which in turn excites 
atom $B$ to the state $|2_B\rangle$, is possible only because of 
$\Gamma_{vc}$ and $\Omega_{vc}$ terms in the master equation (\ref{master}).
Note further that such terms start becoming insignificant as $\hbar\delta$ -
the energy separation between the two excited states increases. Such
interference terms occur in the master equation even when the transition
dipole matrix elements 
$\vec{d}_1$ and $\vec{d}_2$ are orthogonal. Such contributions come from the
second term in Eq.\ (\ref{chi}). In the rest of the paper, we study in detail
various consequences of these interference terms which could be
large and could significantly contribute to the dynamics 
of the system when the atomic separation is smaller than $\lambda$.

Let us consider a geometry where atom $A$ is placed at the origin of a 
Cartesian co-ordinate system and the position vector of atom $B$ 
is $\vec{R}$ (as shown in Fig.\ 2). $\vec{R}$ makes an angle $\theta$ with 
the $z$ axis. Let us assume $\vec{d}_1 = \hat{x} d$ and $\vec{d}_2 = 
\hat{y} d$. All the radiative coupling terms in the master equation
can be written down explicitly as
\begin{eqnarray}
\Gamma_1 = \frac{|d|^2}{\hbar} {\rm Im}~ \chi_{xx} = \frac{3\gamma}{2} (P_i - \sin^2\theta\cos^2\phi~ Q_i)&,&~~ 
\Omega_1 = \frac{|d|^2}{\hbar} {\rm Re}~ \chi_{xx}= \frac{3\gamma}{2} (P_r - \sin^2\theta\cos^2\phi ~Q_r),
\nonumber \\
\Gamma_2 = \frac{|d|^2}{\hbar} {\rm Im}~ \chi_{yy}= \frac{3\gamma}{2} (P_i - \sin^2\theta\sin^2\phi~ Q_i)&,&~~
\Omega_2 = \frac{|d|^2}{\hbar} {\rm Re}~ \chi_{yy}= \frac{3\gamma}{2} (P_r - \sin^2\theta\sin^2\phi~ Q_r)
\\ \nonumber
\Gamma_{vc} = \frac{|d|^2}{\hbar} {\rm Im}~ \chi_{yx}= -\frac{3\gamma}{2} \sin^2\theta\sin\phi\cos\phi~ Q_i&,&~~ 
\Omega_{vc} = \frac{|d|^2}{\hbar} {\rm Re}~ \chi_{yx}= -\frac{3\gamma}{2} \sin^2\theta\sin\phi\cos\phi~ Q_r;
\end{eqnarray}
where $\phi$ is defined as in the Fig.\ 2, and 
\begin{eqnarray}
P_r = \frac{\cos \zeta}{\zeta} - \frac{\sin\zeta}{\zeta^2} -
\frac{\cos\zeta}{\zeta^3}&,&~
Q_r = \frac{\cos \zeta}{\zeta} - 3\frac{\sin\zeta}{\zeta^2} -
3\frac{\cos\zeta}{\zeta^3},
\nonumber \\  
P_i = \frac{\sin \zeta}{\zeta} + \frac{\cos\zeta}{\zeta^2} -
\frac{\sin\zeta}{\zeta^3}&,&~
Q_i = \frac{\sin \zeta}{\zeta} + 3\frac{\cos\zeta}{\zeta^2} -
3\frac{\sin\zeta}{\zeta^3};~\zeta = k_0 R.
\end{eqnarray}
\begin{figure}
\epsfxsize 7cm
\centerline{
\epsfbox{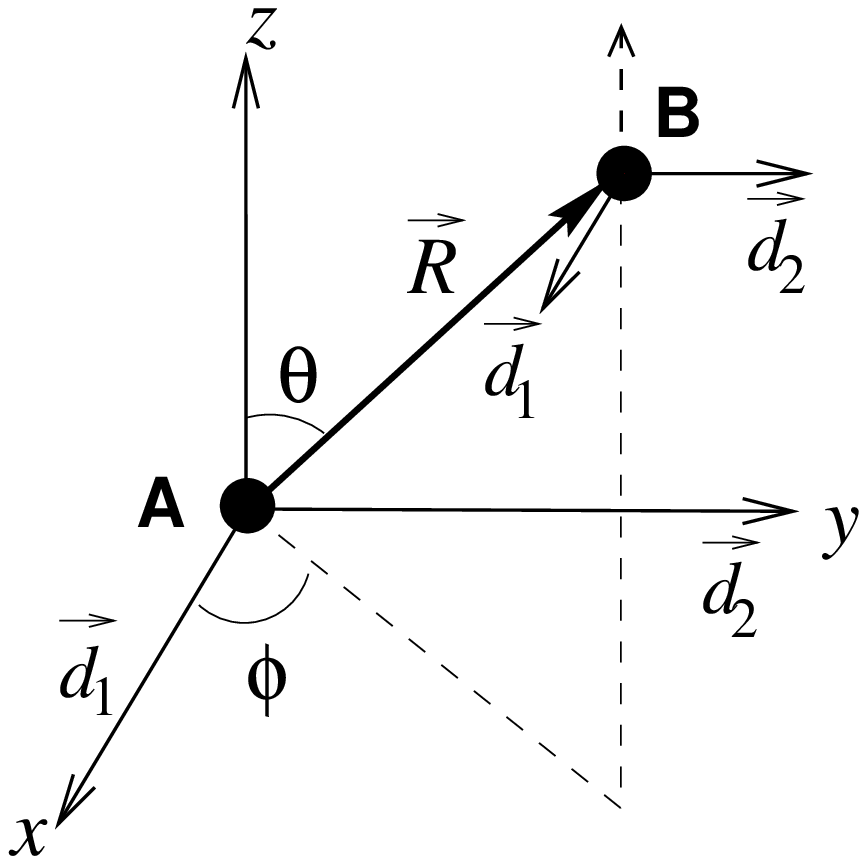}
}
\caption{
The geometry under consideration where the dipole matrix elements 
$\vec{d}_1$ and $\vec{d}_2$ are 
taken to be real and aligned along the $x$ and $y$ directions respectively. 
}
\end{figure}

	In the following, we examine the behavior of the cross coupling 
coefficients responsible for the new coherence effects in different geometries.
In Fig.\ 3 we plot these coefficients as a function of the distance
between the two atoms. In Fig.\ 3(a) we have plotted $\Gamma_1$ and 
$\Gamma_{vc}$, 
and in Fig.\ 3(b) we plot $\Omega_1$ and $\Omega_{vc}$, for comparison.
Clearly, the values of the cross coupling coefficients 
are comparable with the $\Gamma_i$ and $\Omega_i$ values. The value of
$\Omega_{vc}$ becomes significantly large for $R < \lambda/2$. However
for $R\rightarrow 0$, the terms $\Omega_i, \Omega_{vc}$ diverge, 
whereas the terms $\Gamma_i, \Gamma_{vc} \rightarrow 1$.
\begin{figure}
\epsfxsize 9cm
\centerline{
\epsfbox{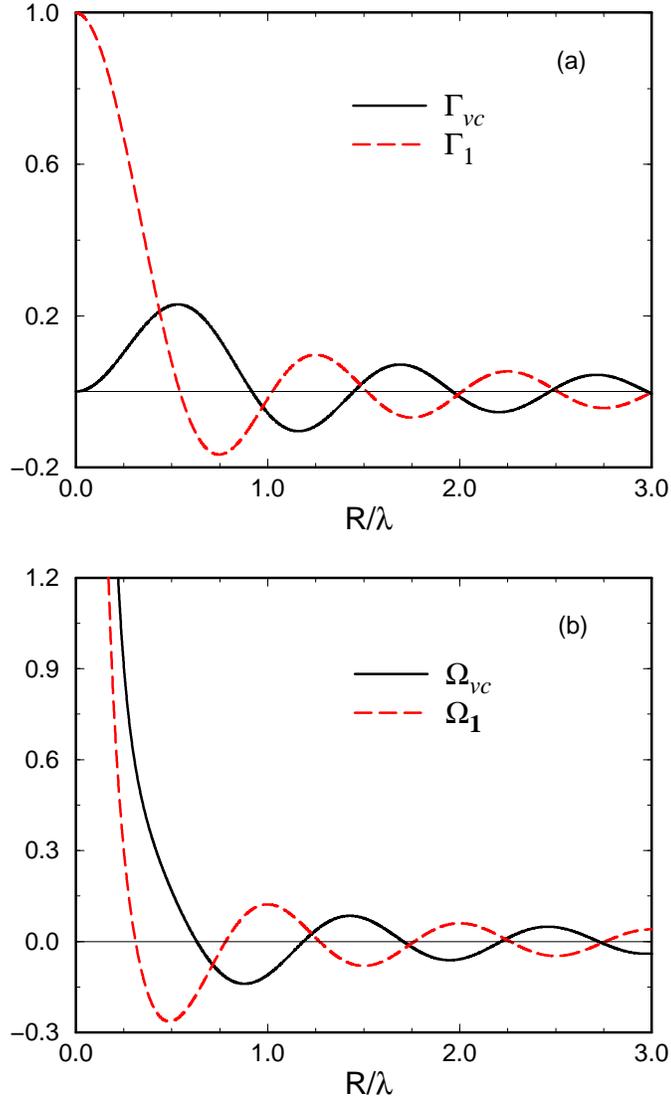}
}
\caption{
Plots of dd-coupling coefficients as a function of the atomic separation.
Here $\theta = \pi/2$, i.e. both the atoms lie on the $xy$-plane and
$\phi = \pi/4$. The new coherence terms $\Gamma_{vc}$, $\Omega_{vc}$ are 
comparable to $\Gamma_i$, $\Omega_i$. All 
the coefficients are scaled with $\gamma$.
}
\end{figure}

Further, in Fig.\ 4 we examine the atomic position dependences of these 
coefficients. We have plotted the coupling coefficients as a function of
$\phi$. Here we have fixed $\theta =\pi/2$; i.e. both the atoms are lying 
in the $xy$-plane.
Again for a comparison, we have plotted $\Gamma_1$ and $\Gamma_{vc}$ in 
Fig.\ 4(a), 
and $\Omega_1$ and $\Omega_{vc}$ in Fig.\ 4(b). We observe the following
special cases:

\noindent
{\em Case} I:
If $\theta = n\pi$, then $\Gamma_{vc} = \Omega_{vc} =0$;
i.e. if $\vec{R}$ is perpendicular to the plane containing $\vec{d}_1$ 
and $\vec{d}_2$, the interference terms in the master equation drop out.

\noindent
{\em Case} II:
When $\phi = n\pi/2$, the coherence terms $\Gamma_{vc} = \Omega_{vc} =0$;
i.e. when the second atom is placed in a position such that
$\vec{R}$ is along one of the dipoles $\vec{d}_1$ or $\vec{d}_2$,
then again the interference terms drop out. Thus the interference effects
in the radiatively coupled systems are sensitive to the geometry.
\begin{figure}
\epsfxsize 9cm
\centerline{
\epsfbox{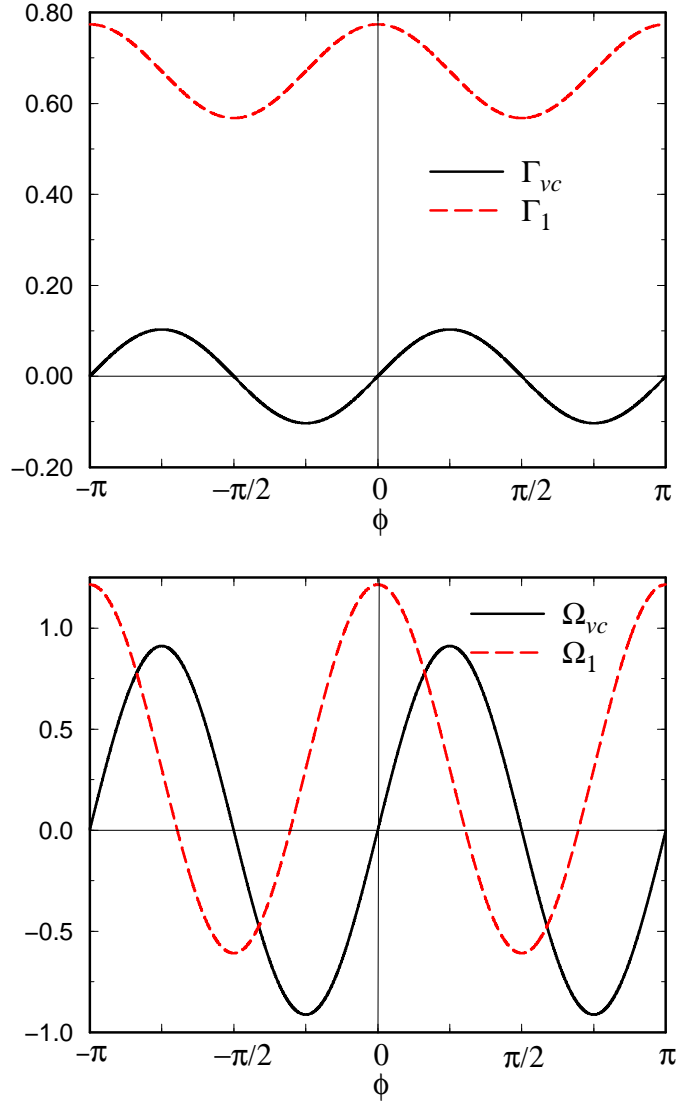}
}
\caption{
The dd-coupling terms as a function of the azimuthal angle $\phi$. Here
$\theta = \pi/2$ and the atomic separation is taken to be $\lambda/4$.
All coefficients are scaled with $\gamma$.
}
\end{figure}

\section{Numerical Results}

In this section we present the numerical results that demonstrate the
effect of the interference terms on the dynamics of the radiatively coupled 
multilevel systems. We use fifth order Runge-Kutta method for the numerical 
solution of the master equation (\ref{master}). For numerical solutions 
we use the initial condition that at $t=0$, the first atom is in excited state 
$|1_A\rangle$ and the second atom is in the ground state $|3_B\rangle$. 

	In Fig.\ 5, we have plotted the density matrix element 
$\rho_{12}^{(A)}\equiv\langle 1_A|\langle 3_B|\rho(t)|2_A\rangle|3_B\rangle$ 
which represents the coherence in the excited states of atom $A$ when atom
$B$ is in ground state $|3_B\rangle$. It is clear from Fig.\ 5 that the 
interference terms $\Gamma_{vc},~ \Omega_{vc}$ in the master equation result
in finite coherence in atom $A$. Otherwise, when $\Gamma_{vc} = \Omega_{vc}
=0$, such coherences vanish. It is 
important to note that this coherence is produced by the radiative coupling
between two atoms even when the dipole matrix elements $\vec{d}_1$ and 
$\vec{d}_2$ are orthogonal.
\begin{figure}
\epsfxsize 12cm
\centerline{
\epsfbox{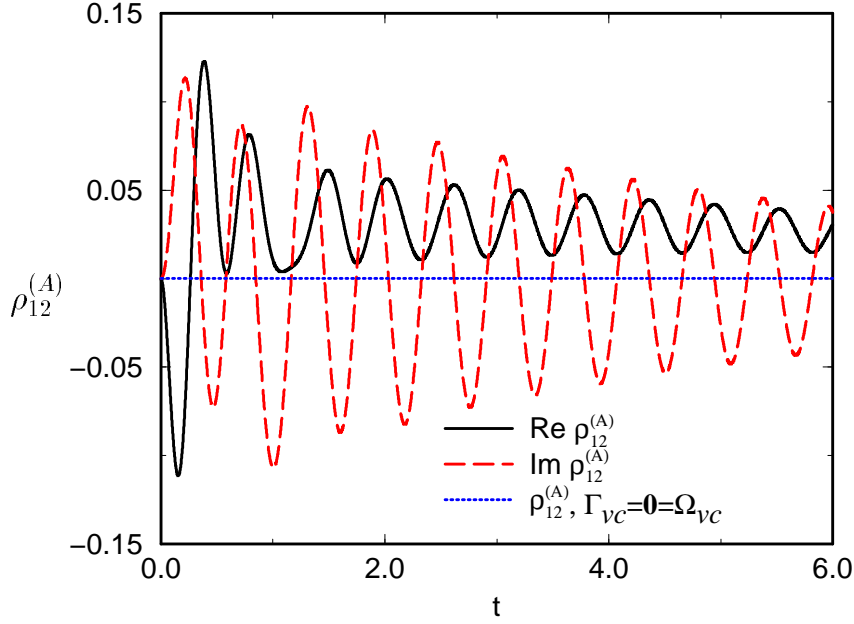}
}
\caption{
Time evolution of the coherence in the excited states of atom $A$
with atom $B$ being in $|3_B\rangle$. The coherence evolves only
for non-zero values of $\Gamma_{vc}$ and $\Omega_{vc}$. Large oscillations
are seen in $\rho_{12}^{(A)}$ which is decided by $\Omega_i,~\Omega_{vc}$, and
$\delta$. The parameters used are $\theta = \pi/2,~\phi=\pi/4,~R=\lambda/2\pi$
and $\delta = 3\gamma$.
}
\end{figure}

	In Figs.\ [6-7], we plot the probabilities that atom $A$ is in 
state $|i_A\rangle$ and atom $B$ is in $|j_B\rangle$, which we denote 
by $p_{i;j} \equiv \langle i_A|\langle j_B|
\rho(t)|i_A\rangle|j_B\rangle$. In Fig.\ 6(a), we present $p_{3;2}(t)$ 
that represents the simultaneous probability of atom $A$ being 
de-excited to state $|3_A\rangle$ and atom $B$ being excited to the state
$|2_B\rangle$. Fig.\ 6(b) is the plot of $p_{2;3}(t)$ that represents the
probability that the atom $A$ is excited to state $|2_A\rangle$ with atom $B$
being in $|3_B\rangle$. Obviously both 
$p_{3;2}$ and $p_{2;3}$ become zero if $\Gamma_{vc}=\Omega_{vc}=0$.
It is observed that smaller the atomic separation larger is the excitation 
probability. For atomic separation $R = \lambda /2\pi$, the excitation
probabilities are very large, e.g. more than $25\%$ of the population
in atom $B$ could be excited to state $|2_B\rangle$ at $t\sim 0.3/\gamma$
(Fig.\ 6(a)) and, similarly in atom $A$, $\sim 18.5\%$ population could be 
excited to state $|2_A\rangle$ at $t\sim 0.5/\gamma$. Thus significant 
amount of energy transfer can take place between the states $|1_A\rangle$
and $|2_B\rangle$, though the corresponding transition dipoles are 
orthogonal to each other. 
Note that the initial evolution of $p_{2;3}$ is much slower compared to 
the evolution of $p_{3;2}$. This can be understood as
follows: The excitation of atom $B$ to the state $|2_B\rangle$ can be 
caused by a single photon transfer from $A$ to $B$ [the process 
$|1_A,3_B\rangle \rightarrow |3_A,2_B\rangle$], whereas the excitation of 
atom $A$ to the state $|2_A\rangle$ occurs only through atom $B$ 
and this involves a net transfer of two photons [processes 
$|1_A,3_B\rangle\rightarrow |3_A,2_B\rangle\rightarrow |2_A,3_B\rangle$ or 
$|1_A,3_B\rangle\rightarrow|3_A,1_B\rangle\rightarrow |2_A,3_B\rangle$].
The oscillatory character of $p_{3;2}$ and $p_{2;3}$ comes from non-vanishing
$\delta$ and from the dd-coupling co-efficients $\Omega_i$ and $\Omega_{vc}$. The 
excitation probabilities are seen to be larger for degenerate excited states 
($\delta = 0$) compared to that with finite separation between the 
excited states. For very large 
$\delta$ ($>> \gamma$), this interference effect disappears.

\begin{figure}
\epsfxsize 10cm
\centerline{
\epsfbox{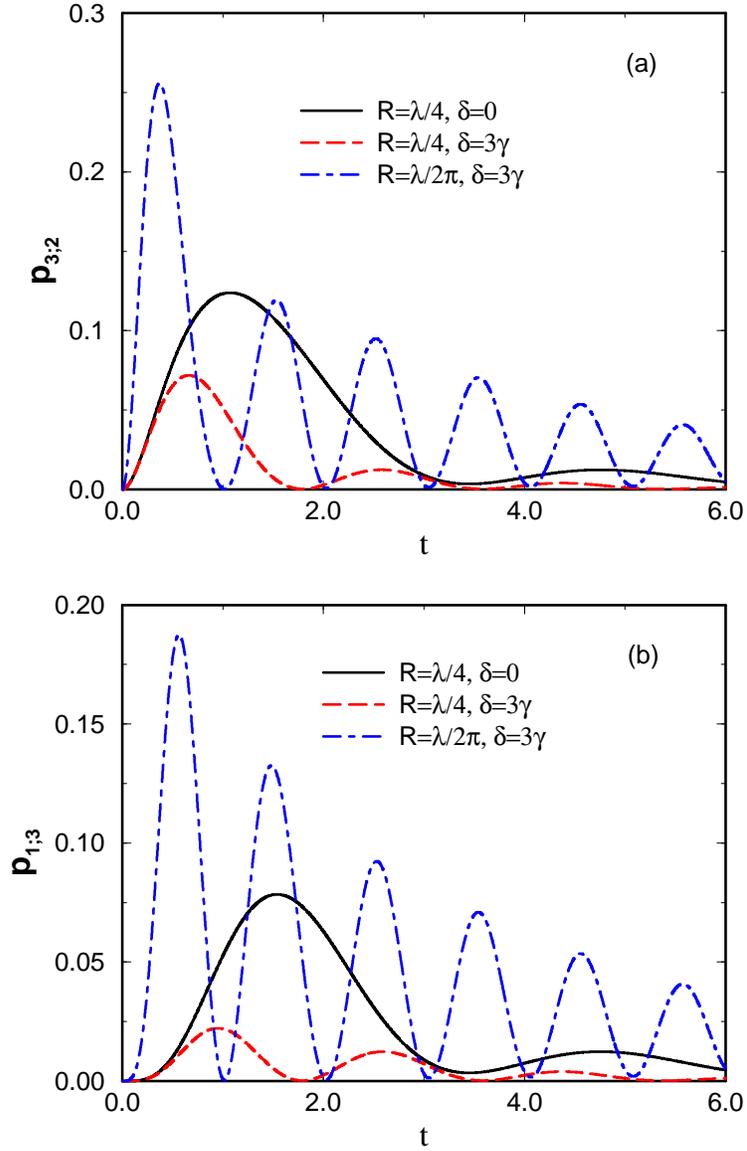}
}
\caption{
The time evolution of (a) $p_{3;2}$ and (b) $p_{2;3}$ are plotted for 
different parameters with the initial condition that atom $A$ is in
state $|1_A\rangle$ and atom $B$ in $|3_B\rangle$. In both cases 
$\theta=\pi/2$ and $\phi=\pi/4$. The values of other parameters are shown as
legends. 
}
\end{figure}

	In Fig.\ 7, we present a comparative study of the probability that
atoms remain in their initial states, i.e. $p_{1;3}$, in the presence and 
absence of the dd-coupling terms.
The probability of atom $A$ staying in $|1_A\rangle$
decays exponentially in the absence of atom $B$. However, in the presence 
of the second atom, the nature of its decay is significantly
modified - large oscillations are seen in $p_{1;3}$
in the presence of the new coherence terms, which is evident from Fig.\ 7. 
The origin of this oscillation is attributed to the large values of $\Omega_{vc}$. 
\begin{figure}
\epsfxsize 10cm
\centerline{
\epsfbox{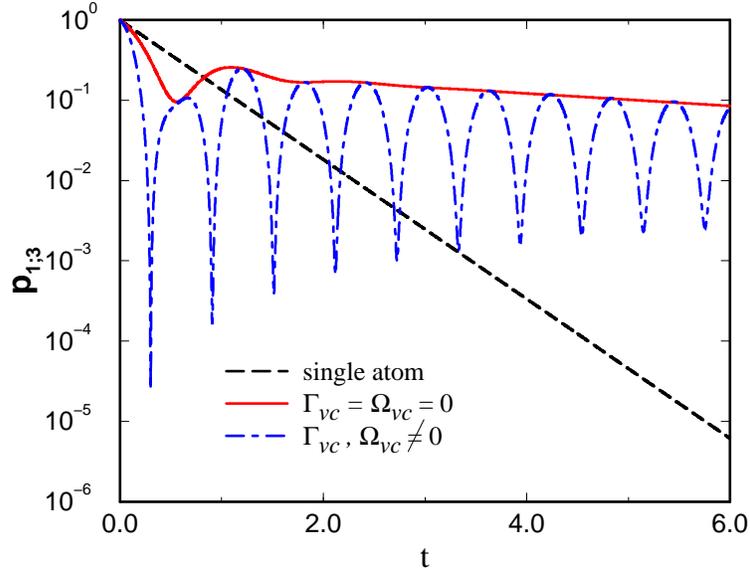}
}
\caption{
Plot of the probability that both atoms remain in their initial states.
This probability is plotted on a logarithmic scale as a function of time
in linear scale. The parameters used are $\theta =\pi/2, ~\phi=\pi/4,~
R=\lambda/4$ and $\delta=0$.
}
\end{figure}

\section{Two $V$-systems with magnetic sub-levels 
in the presence of a magnetic field}

In this section we consider the new coherence effects in two $V$-systems with 
$m$-degenerate magnetic sub-levels as excited states. The system could be,
for example, a $^{40}$Ca system - where $4~^1P_1$ degenerate sublevels would
correspond to the excited states $|1_\mu\rangle \equiv |j=1, m=1\rangle$ and
$|2_\mu\rangle \equiv |j=1, m=-1\rangle$, and the $4~^1S_0$ state would 
correspond to the ground state $|3_\mu\rangle$.
In this case the dipole matrix elements $\vec{d}_1$ and $\vec{d}_2$ are 
complex and orthogonal to each other, and are given by
\begin{equation}
\vec{d}_1 = -d\hat{\epsilon}_-, ~ \vec{d}_2= d \hat{\epsilon}_+;~ 
\hat{\epsilon}_\pm = \frac{\hat{x}\pm i\hat{y}}{\sqrt{2}}, 
\end{equation}
where $d$ is the reduced dipole matrix element. The magnetic field 
produces a Zeeman splitting $\delta$ and 
fixes the quantization axis ($z$ axis in our case). The geometry 
can be taken to be the same as in Fig.\ 2. However, 
in the present case, $\vec{d}_1$ and $\vec{d}_2$ being complex dipoles,
they are not fixed along the real axes unlike in Fig.\ 2. Using Eq.\ (\ref{coeff}),
the dd coupling
coefficients for this scheme can be obtained
\begin{eqnarray}
\Gamma_1&=& \Gamma_2 = \frac{|d|^2}{2\hbar} {\rm Im}~(\chi_{xx}+\chi_{yy})\equiv
\frac{3\gamma}{4}(2P_i -\sin^2\theta Q_i), 
\nonumber \\
\Omega_1 &=& \Omega_2 = \frac{|d|^2}{2\hbar} {\rm Re}~(\chi_{xx}+\chi_{yy})
\equiv \frac{3\gamma}{4}(2P_r -\sin^2\theta Q_r),
\nonumber \\
\Gamma_{vc} &=& -\frac{|d|^2}{2\hbar} \left[{\rm Im}~(\chi_{xx}-\chi_{yy})
+ i~ {\rm Im}~(\chi_{xy}+\chi_{yx})\right] 
\equiv \frac{3\gamma}{4} \sin^2\theta e^{2i\phi}Q_i,
\nonumber \\
~~\Omega_{vc} &=& -\frac{|d|^2}{2\hbar} \left[{\rm Re}~(\chi_{xx}-\chi_{yy})
+ i~ {\rm Re}~(\chi_{xy}+\chi_{yx})\right] 
\equiv \frac{3\gamma}{4} \sin^2\theta e^{2i\phi}Q_r.
\label{coeff-complex}
\end{eqnarray}
The $P$s' and $Q$s' are as defined in Eq.\ (26). In deriving 
(\ref{coeff-complex}),
we have used the fact that $\chi_{xy} = \chi_{yx}$. It may be noted that
$\Gamma_i$ and $\Omega_i$ are real, and are independent of the azimuthal
angle, whereas $\Gamma_{vc}$ and $\Omega_{vc}$ are complex and are functions
of $\phi$. For $\theta = n\pi$, the coherence terms disappear in 
Eq.\ (\ref{master}). Thus if $\vec{R}$ is perpendicular to the plane containing
both the dipoles, i.e. both atoms lie on the quantization axis ($z$-axis), the 
coherence effects vanish. 

The solutions of the master equation can be recalculated using the above
coefficients and the analog of all the results presented in Sec.IV can be produced
for the present system. For completeness, we present the numerical plot that
shows the excitation probability $p_{3;2}$ with the initial condition 
$p_{1;3}(t=0) = 1$. The time evolution of $p_{3;2}$ is similar to the one in 
the case of real dipoles (cf. Fig.\ 6(a)). 

\begin{figure}
\epsfxsize 10cm
\centerline{
\epsfbox{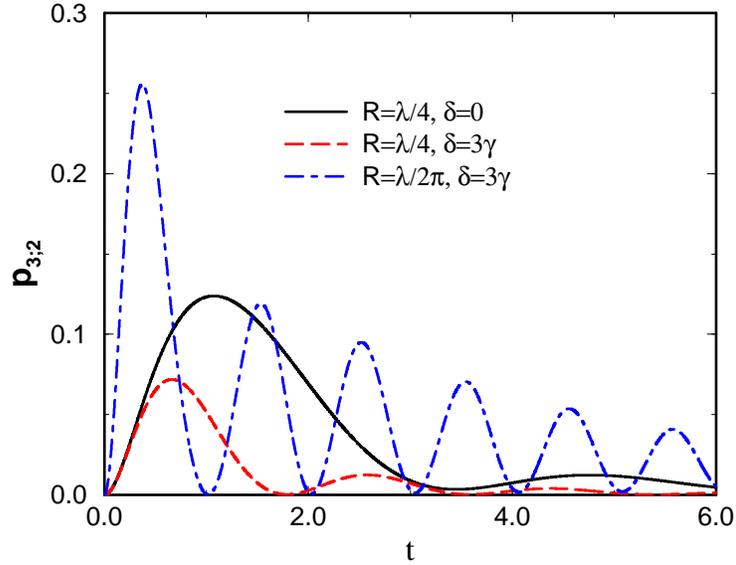}
}
\caption{
The time evolution of the probability $p_{3;2}$ when the dipoles 
$\vec{d}_1$ and $\vec{d}_2$ are 
complex. All the parameters are same as in Fig.\ 6. 
}
\end{figure}
It may further be noted that $p_{3;2}$ is independent of $\phi$
though $\Gamma_{vc}$ and $\Omega_{vc}$ are functions of $\phi$. This
is because $p_{3;2}$ is a function of the absolute values of $\Gamma_{vc}$
and $\Omega_{vc}$. - which can be shown from (\ref{den-elmnt}) and 
(\ref{coeff-complex}), to the lowest order in $\Gamma_{vc}$ and 
$\Omega_{vc}$, 
\begin{equation}
p_{3;2} \approx
4\left( |\Gamma_{vc}|^2 +|\Omega_{vc}|^2 \right) 
\left( \frac{\sin^2(\delta/2)}{\delta^2} \right).
\end{equation}

	We now discuss how the new coherence effect 
can be monitored experimentally for the above mentioned system. The dipole 
transitions $|1_\mu\rangle 
\leftrightarrow |3_\mu\rangle$, in the system described above, involve photons 
having $\sigma_+$ polarization.
Thus the emission from $|1_\mu\rangle \rightarrow |3_\mu\rangle$ does not 
contain any 
field component in $\sigma_-$ polarization. On the other hand, the emission
from $|2_\mu\rangle \rightarrow |3_\mu\rangle$ would contain $\sigma_-$
component. Thus the signal that one has to look for is - the intensity of
the emitted photon from $|2_\mu\rangle$ levels in $\sigma_-$ polarization, which 
would a be measure of the total excitation probability to $|2_\mu\rangle$ states
and hence would confirm the occurrence of VIC. Another possibility to
probe the population in $|2_\mu\rangle$ will be to excite it with a circularly 
polarized radiation to a fourth state $6~^1S_0$ and to monitor the 
fluorescence from $6~^1S_0$. 

\section{Conclusions}
	In conclusion, we have shown that the radiative coupling between
multilevel atoms with near-degenerate transitions can produce new interference
effects which are especially important when the distance between two dipoles is
less than a wavelength. We have demonstrated this possibility by 
considering two identical $V$-systems such that the pair of transition dipole
matrix elements in each system are orthogonal to each other in both the atoms. 
Such 
interference effects are especially significant in the energy transfer studies. 
The choice of orthogonal dipole matrix elements enables us to specially 
isolate the effects of the vacuum induced coherences in the radiative
coupling between multilevel atoms with nearly degenerate transitions. We have 
presented detailed numerical results to bring out the role of multi-atom
multilevel interference effects.

\end{document}